\documentclass[twocolumn,showpacs,preprintnumbers,amsmath,amssymb,prl]{revtex4-1}
\usepackage{graphicx}% Include figure files
\usepackage{dcolumn}% Align table columns on decimal point
\usepackage{bm}% bold math

\newcommand{\R}{{\bm{R}}}

\renewcommand{\d}{{\bm{\delta}}}

\renewcommand{\k}{{\bm{k}}}

\renewcommand{\r}{{\bm{r}}}

\def\gsim{\lower.35em\hbox{$\stackrel{\textstyle>}{\textstyle\sim}$}}

\begin{document}
%\draft

%\title{Measurable lattice effects on the charge and current response in graphene}
%\title{Lattice effects on the charge and magnetic response in graphene}
\title{Measurable lattice effects on the charge and magnetic response in graphene}
\author{G. G\'omez-Santos$^1$ and T. Stauber$^{1,2}$}

\affiliation{$^1$Departamento de F\'{\i}sica de la Materia Condensada  and Instituto Nicol\'as Cabrera, Universidad Aut\'onoma de Madrid, E-28049 Madrid, Spain}
\affiliation{$^2$Centro de F\'{\i}sica  e  Departamento de
F\'{\i}sica, Universidade do Minho, P-4710-057, Braga, Portugal}
%\date{\today}

\begin{abstract}
The simplest tight-binding model is used to study lattice effects on
two properties of doped graphene: i) magnetic orbital
susceptibility and ii) regular Friedel oscillations, both suppressed in the
usual Dirac cone approximation. i) An exact expression for the
tight-binding magnetic susceptibility is obtained, leading to orbital
paramagnetism in graphene for a wide range of doping levels which is
relevant when compared with other contributions. ii) Friedel
oscillations in the coarse-grained charge response are considered
numerically and analytically and an explicit expression for the response
to lowest order in lattice effects is presented, showing the
restoration of regular {\it 2d} behavior, but with strong sixfold
anisotropy.
\end{abstract}

%\maketitle

  
\maketitle 

%%%%%%%%%%%%%%%%%%%%%%%%%%%%%%%%%%%%%%%%%%%%%%%%%%%%%%%%%%%%%
%  SECTION INTRODUCTION
%%%%%%%%%%%%%%%%%%%%%%%%%%%%%%%%%%%%%%%%%%%%%%%%%%%%%%%%%%%%%

{\it Introduction.} The recent experimental realization
\cite{Novos04} of graphene, the single layer honeycomb lattice
of carbon atoms that forms graphite, has unleashed an explosion of
activity.  High expectations have been put on profiting from its
peculiar electronic, mechanical, optical (and perhaps
magnetic) properties, when tailored at the
nanoscale \cite{Geim09}. The existence of linearly dispersing
bands around two nodal points ({\it massless} Dirac fermions with
velocity $v\sim 10^6 \; m/s$) form the basis of graphene's most
notable electronic properties \cite{Castro09}.

Many theoretical studies of graphene are done within scaling limit or
Dirac cone approximation, that is, assuming strictly linear energy
dispersion around the nodal points. Although this approach is
successful in explaining many experimental facts, it has limitations
too. Obvious examples are provided by magnitudes for which the
Dirac cone approximation provides a null result. In this paper we are
concerned with two such magnitudes: i) the magnetic susceptibility and
ii) regular Friedel oscillations, both rendered zero at finite doping
in the scaling limit.

i) Strong and peculiar diamagnetism characterizes graphene, as first
discussed by McClure to explain graphite, Nature's best diamagnet.  He
found that, for the two-dimensional Dirac model, the diamagnetic
susceptibility was given by a delta-function of the chemical
potential \cite{McClure56,*Koshino09,*Principi09}. This result implies that there is no
magnetic response when the chemical potential is shifted from the
neutrality point. This is in clear contrast with recent experimental
findings of paramagnetism in graphene \cite{Geim10}.

Here we will show that lattice effects, neglected in the scaling
limit, render finite and sizable the magnetic response.  We employ
the formalism of Fukuyama \cite{Fukuyama71}, whose original formula was
first applied to graphite \cite{Sharma74} and subsequently to graphite
intercalated compounds \cite{Safran79,Blinowski84}, which is here
extended by an additional term required to provide the exact
susceptibility for a general tight-binding model.  The magnetic
response for arbitrary chemical potential is obtained, finding orbital
paramagnetism (OP) over a wide range of fillings.  Its value close to
the neutrality point is compared with other sources (core
diamagnetism, spin paramagnetism and interaction's induced orbital
paramagnetism) and shown to be a relevant contribution.

ii) Graphene's charge response around localized perturbations is also
peculiar \cite{Cheianov06,Wunsch06,Mariani07,*Brey07,*Bena08}. While ordinary
{\it 2d} systems show the familiar Friedel $2k_F$ oscillations
decaying as $1/r^2$, graphene's {\it coarse-grained} response in the
scaling limit does so but with an additional power. Graphene's lack of
regular {\it 2d} Friedel oscillations is linked to isospin (or chiral)
conservation and thus provides the possibility of direct
observation of the nature of graphene's
excitations \cite{Brihuega08}.

Here we show, numerically and analytically, that lattice effects restore
the standard {\it 2d} behavior. An explicit expression for the charge
response is obtained to lowest order in lattice effects, exhibiting
the usual $1/r^2$ decay and a pronounced sixfold anisotropy, with
maxima along the bond's directions.  

One might ask why we treat two at first sight such distinct topics on  the same
footing. The reason is that within the Dirac cone approximation the static
transverse current-current as well as the charge-charge correlation function
yield $\chi(q)=a+bq^2+...$ with $b\sim\delta(E_F)$ and $E_F$ the Fermi level.
 Lattice contributions to the response, given by $\chi^{lattice}\sim  
q^2$ for finite filling factor, are thus suppressed in the same  
peculiar way.

%%%%%%%%%%%%%%%%%%%%%%%%%%%%%%%%%%%%%%%%%%%%%%%%%%%%%%%%%%%%%
%  SECTION HAMILTONIAN. MAGNETIC RESPONSE
%%%%%%%%%%%%%%%%%%%%%%%%%%%%%%%%%%%%%%%%%%%%%%%%%%%%%%%%%%%%%
 
{\it Tight-binding model and Dirac cone approximation.} We describe
graphene by the simplest tight-binding Hamiltonian
$H=-t\sum_{\R,\d}a_\R^\dagger b_{\R+\d}^{}+H.c.$ with hopping
amplitude $-t$ between nearest neighbor atoms in A and B sublattices
joined by vectors $\bm \delta_1 = (0,a) $ and its $\pm 120^o$-rotated
versions $ \bm \delta_{2,3}$.  The spectrum is $E^{\pm}(\bm k)= \pm|t
S(\bm k)|$, where $S(\bm k) = \sum_{i} \exp(i \bm k \cdot \bm
\delta_i) $. It develops a well-known linear dispersion $E^{\pm}(\bm
k)= \pm \hbar v k' $ with $\bm k' = \bm k - \bm K_{1,2}$ around two
points in the Brillouin zone, $\bm K_{1,2} = \pm (\tfrac{4 \pi}{3
  \sqrt{3} a},0)$.

For the orbital magnetic response, the Dirac cone approximation leads
to a diamagnetic susceptibility depending on Fermi level $E_F$
as \cite{McClure56,Koshino09,Principi09} 
\begin{equation}\label{deltadiag}
\chi_{orb}^{Dirac} = -\mu_o \frac{g_s g_v}{6 \pi} e^2 v^2 \delta(E_F) 
,\end{equation}
with vacuum permeability $\mu_o$ (SI units), spin  and valley  degeneracies
$g_s = g_v = 2$, and unit charge $e$. The
peculiar relation, known to be at the basis of graphite's large
diamagnetism \cite{Sharma74}, implies that doped graphene shows no
magnetic orbital response at finite doping. 

Also the charge response to a local impurity $V=u\delta(\r)$ shows
peculiar behavior since the first $q$-derivative of its susceptibility is
continuous at $q=2 k_F$. This results in an anomalous decay of the Friedel
oscillations which in terms of the electronic carrier density $\rho_e$
reads \cite{Cheianov06}
\begin{equation}\label{FOAnalyticDirac}
\frac{\delta\rho(\bm r)}{\rho_e} = \frac{u\rho_e}{E_F}
 \frac{\cos(2 k_F r)}{(2 k_F r)^3}\;,\; k_Fr\gg1\;.
\end{equation}
In what follows, we will show that both results are substantially
altered when lattice effects are included.

%%%%%%%%%%%%%%%%%%%%%%%%%%%%%%%%%%%%%%%%%%%%%%%%%%%%%%%%%%%%%
%  SUBSECTION susceptibility
%%%%%%%%%%%%%%%%%%%%%%%%%%%%%%%%%%%%%%%%%%%%%%%%%%%%%%%%%%%%%

{\it Magnetic response.} Fukuyama \cite{Fukuyama71} has provided a
convenient expression for the orbital magnetic susceptibility in
non-interacting systems. The formula is exact only for a Hamiltonian
of the canonical form $ H = \frac{\bm P^2}{2m} + V(\bm R) $. Here we
adapt Fukuyama's procedure to obtain the exact orbital response of a
tight-binding Hamiltonian. Employing the current operator of the
tight-binding model given in Ref. \cite{Stauber10}, linear
response theory yields the following expression for the orbital
susceptibility \cite{supplmaterial}:
\begin{align}\label{chi}
&\chi_{orb} = - \mu_o \frac{e^2}{\hbar^2}
 \frac{g_s}{ 2\pi} \text{Im} \int \!\! dE \, n_F(E)\, \frac{1}{A} \sum_{\bm k} \\
 & \text{Tr}\{ 
\hat{\gamma}^{x}%_{\bm k}
\hat{\cal G}%_{\bm k}
\hat{\gamma}^{y}%_{\bm k} 
\hat{\cal G}%_{\bm k}   
\hat{\gamma}^{x}%_{\bm k}
\hat{\cal G}%_{\bm k}
\hat{\gamma}^{y}%_{\bm k} 
\hat{\cal G}%_{\bm k} 
+ 
\frac{1}{2}(
\hat{\cal G}%_{\bm k}
\hat{\gamma}^{x}%_{\bm k} 
\hat{\cal G}%_{\bm k}
\hat{\gamma}^{y}%_{\bm k} 
+
\hat{\cal G}%_{\bm k}
\hat{\gamma}^{y}%_{\bm k} 
\hat{\cal G}%_{\bm k}
\hat{\gamma}^{x}%_{\bm k}) ) 
)\hat{\cal G}%_{\bm k} 
\frac{\partial\hat{\gamma}^{y}%_{\bm k}
 }{\partial k_x}
\} 
\nonumber
\end{align}
with the 
$2\!\times\!2$ matrix
$\hat{\cal G}_{\bm k}(E) = (E +i0^+ - \hat{H}_\k)^{-1} $, $\hat{ \bm
\gamma}_{\bm k} = \bm \nabla_{\bm k} \hat{H}_\k$, and   
$\hat{H}_\k$ given by
\begin{equation}
\hat{H}_\k=
\begin{pmatrix}
0& -t \; S(\bm k) \\ -t \; (S(\bm k))^*& 0 
\end{pmatrix}
.\end{equation}

This expression for $\chi_{orb}$ transcends the model of the initial
Hamiltonian, and turns out to be correct for any tight-binding system.
Eq. (\ref{chi}) coincides with Fukuyama's original result
\cite{Fukuyama71} except for the second term. 
The difference stems from the standard isotropic $\bm P$-dependence in
Fukuyama's $H = \frac{\bm P^2}{2m} + V(\bm R) $, where one has
$\frac{\partial\hat{\gamma}^{y}}{\partial k_x} = 0 $.  Such
cancellation does not apply in a generic tight-binding case.

\begin{figure} % Requires \usepackage{graphicx} 
\includegraphics[clip,width=8cm]{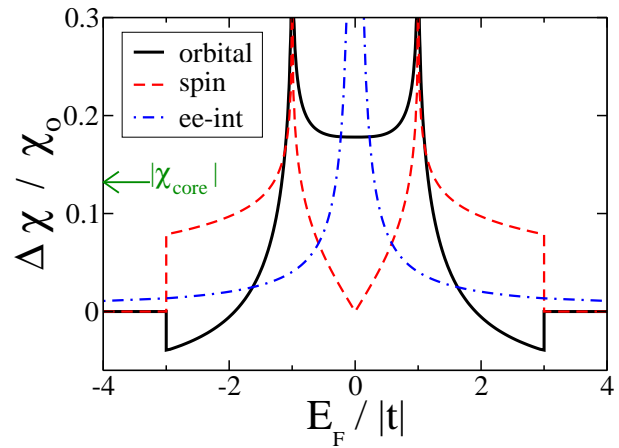} \\ 
\caption{ Continuous line: lattice contribution to the orbital magnetic 
susceptibility $(\Delta \chi_{orb})$ in units of 
$\chi_{o} = \mu_o \hbar^{-2} e^2 |t| a^2$. Dashed line: Pauli's spin paramagnetic
contribution. Dashed-dotted line: orbital magnetic susceptibility from
Coulomb electron-electron interactions (Eq. (\ref{chiint})). The arrow marks an
estimate of the absolute value of the diamagnetic contribution from core 
electrons \cite{DiSalvo79}.}
\label{figorbital}
\end{figure}

The above formula has been applied to numerically calculate the orbital susceptibility in
graphene as a function of Fermi level $E_F$.  The most prominent feature is, of course,
the delta function at the band center of Eq. (\ref{deltadiag}), which comes with the
known analytical value.  We can thus extract the lattice contribution from the numerical
results by writing 
\begin{equation}\label{Deltachi} \chi_{orb} = -\mu_o \frac{g_s
g_v}{6 \pi} e^2 v^2 \delta(E_F) + \Delta \chi_{orb} 
.\end{equation}
The calculated
lattice contribution $\Delta \chi_{orb}$ is  plotted in Fig. \ref{figorbital} in units of
$\chi_{o} = \mu_o \hbar^{-2} e^2 |t| a^2$. The lattice origin of $ \Delta \chi_{orb}$
becomes evident if one artificially  sets  $ a\rightarrow 0 $ 
and $t\to\infty$ while $at\propto v=$ constant
(scaling limit): then $
\Delta \chi_{orb}\rightarrow 0$, leaving the  Dirac cone result (Eq. (\ref{deltadiag}))
as the sole response.

{\it Discussion.} The orbital response is usually associated with
diamagnetism, so a noteworthy aspect of the lattice contribution in
Fig. \ref{figorbital} is its {\em paramagnetic} character over most of
the band, even diverging at the van Hove points \cite{Vignale91}.   
From Eq. (\ref{chi}), one
finds the following sum rule: $\int \! dE_F \; \chi_{orb}(E_F)= 0$. The
existence of OP is thus a necessary consequence to cancel the large diamagnetic
contribution of the scaling limit at the band center, see Eq. (\ref{Deltachi}). Only at the band edges Landau diamagnetism emerges, as expected.

Now, we compare lattice's OP with other contributions in the region
$E_F \sim 0$, relevant for gate-doped graphene.  Within our
non-interacting model, the only remaining magnetic contribution is
Pauli's spin paramagnetism, given by $\chi_{spin} / \mu_o= 2 \mu_B^2
\rho(E_F)$ where $\mu_B$ is the Bohr magneton and $\rho(E_F)$ is the
density of states (per spin). This spin contribution is plotted in
Fig. \ref{figorbital}, where it is seen that it cannot compete with
the dominant orbital contribution for low carrier densities $\rho_e$.

Core electrons, not  considered in our   $\text{pp}\pi$ band Hamiltonian,  are
another source of (dia)magnetic response.  The estimate of  Ref.
\cite{DiSalvo79}, $\chi_{core}\sim -4.8\times 10^{-6}\text{emu/mol}$,
translates into $\chi_{core}\sim -0.13\,\chi_{o} $, a value marked with an
arrow in the scale of Fig. \ref{figorbital}. Again, the orbital contribution for low $\rho_e$ is comparable or greater than this estimate of core
diamagnetism.    

Electron-electron interaction is the major ingredient left out in our model.
Recently, its effect on the magnetic response has been calculated to
first order within the Dirac cone approximation \cite{Polini10}. Two scenarios 
have been considered: Thomas-Fermi screened Coulomb interaction and contact
(Hubbard-like) interaction, both leading to OP at finite
doping. For the Coulomb case,  the interaction's contribution to the
susceptibility can be written as \cite{Polini10}
\begin{equation}\label{chiint}
\chi_{ee} =  \chi_o \frac{C}{E_F/|t|}\;,
\end{equation}
with an interaction dependent constant  $C\sim 0.04$, suitable  for graphene
over $\text{SiO}_2$.  $\chi_{ee}$ is compared with the lattice contribution in
Fig. \ref{figorbital}. Graphene's poor screening causes the divergence of
$\chi_{ee}$ when $E_F\rightarrow 0$,  making this contribution dominant. But
even in this unfavorable case, the lattice contribution is not negligible. For
instance, $\Delta \chi_{orb} \gtrsim \chi_{ee} $ for doping levels
$n\!\gtrsim\!3.7 \times 10^{13} \;\text{cm}^{-2} $, and for a typical doping $n
\sim 4 \times 10^{12} \;\text{cm}^{-2} $ one has  $\chi_{ee} \sim 3 \Delta
\chi_{orb}  $.  

Finally, if  (by some external means) screening were truly effective so that
interactions could be described by a contact term $v_o$,  Ref.
\cite{Polini10}  provides the following expression for the interaction's
promoted orbital  paramagnetism:
\begin{equation}\label{chiintU}
\chi_{ee} = g_s g_v \mu_o e^2 \hbar^{-2}  \frac{13}{256 \pi^2} v_o \sim 
0.027 \;\chi_o\;  \frac{U}{|t|}
,\end{equation}
where we have written the interaction as $ v_o = UA_c/2$, with a
Hubbard-like energy $U$, and area per unit cell $A_c =
\tfrac{3\sqrt{3}a^2}{2} $.  In order to compare this contribution with
that of the lattice, we calculate the size of $U$ required for
$\chi_{ee}$ of Eq. (\ref{chiintU}) to match the lattice contribution
around $E_F \sim 0 $. The answer turns out to be $U\sim 7 |t|$, a
value substantially larger than current estimates for graphene. This
implies that, in any reasonable scenario of contact interactions, the
lattice orbital contribution to paramagnetism would be the key
player in the magnetic response of doped graphene.

Let us close with a remark on the effect of next-nearest neighbor hopping, temperature
and disorder. As was already noted in Ref. \cite{Blinowski84}, $t'\approx0.1t$ leads to a
considerable electron-hole asymmetry in the magnetic response; the above qualitative
discussion on the relevance of the several contributions, though, is not altered.
Temperature and disorder\cite{Ando,*Nakamura} broadens the diamagnetic delta-peak, such
that lattice effects gradually lose relevance at a given temperature or disorder when the
chemical potential decreases to zero. 

%%%%%%%%%%%%%%%%%%%%%%%%%%%%%%%%%%%%%%%%%%%%%%%%%%%%%%%%%%%%%
%  SECTION Charge Response. Friedel oscillations
%%%%%%%%%%%%%%%%%%%%%%%%%%%%%%%%%%%%%%%%%%%%%%%%%%%%%%%%%%%%% 
{\it Charge Response.} The linear, static charge response of graphene is
  given by
\begin{equation}\label{chirho}
\Pi(\bm q) = \frac{1}{A}\!\!\!\sum_{\bm k_1, s,s'=\pm}\!\!\! f_{s \cdot s'}(\bm
k_1,\bm k_2) \frac{n_F(E^s(\bm k_1)) - n_F(E^{s'}(\bm k_2))}{E^s(\bm
  k_1) - E^{s'}(\bm k_2) }\;,
\end{equation}
%
%with system's area $\mathcal S$,Fermi function $n_F(E)$ 
with $\bm k_1 = \bm k_2 + \bm q$, 
and the prefactor $f_{\pm}$
\begin{equation}\label{prefactor}
f_{\pm}(\bm k_1,\bm k_2) = \frac{1}{2} \pm \frac{1}{2}
\text{Re}\left( \frac{S(\bm k_1)}{|S(\bm k_1)|} \frac{S^*(\bm k_2)}{|S(\bm k_2)|}  \right)\;. 
\end{equation}

Friedel oscillations are caused by intraband transitions 
($+$-sign in Eq. (\ref{prefactor})) with $q/2 \sim  k_F$, 
the Fermi wavevector (measured from the  Dirac
point). To understand graphene's
peculiarity, let us set 
$f_+ = 1 $ in Eq. (\ref{chirho}), and call the associated {\em Lindhard}-like response
$\tilde{\Pi} $. Then, the dominant singularity in $\tilde{\Pi}$,
corresponding to transitions across the Fermi surface, leads to the
prototypical {\it 2d} square-root behavior
\begin{equation}\label{chirhofake}
\tilde{\Pi}(q) \sim  \frac{g_s g_v \sqrt{k_F}}{2 \pi \hbar v} \sqrt{q - 2 k_F} \; \Theta(q - 2 k_F)
,\end{equation}
where $\Theta$ is Heaviside's function and we have ignored any distortion of the
 isotropic Fermi surface around the two Dirac points.
 Within the Dirac cone approximation, the prefactor $f_+$ in
Eq. (\ref{chirhofake})  crucially vanishes for states on
opposite sides of the Fermi surface. The physical interpretation is 
well known \cite{Cheianov06}: the involved
states have opposite chirality and cannot be coupled by a perturbation
 diagonal in sublattice index. Nevertheless,
this {\it exact} cancellation of $f_+$ holds true only in the scaling
limit $k_F a \rightarrow 0$, and a {\em finite} value of $k_F a $
renders $f_+$ finite, something we generically label as {\em lattice
effect}.

To see if this square root behavior is present also for the true 
prefactor $f_+$ as given in Eq. (\ref{chirho}), 
we numerically analyze the response derivatives which 
we conveniently write as
\begin{align}\label{numer}
\frac{\partial\Pi(\bm q)} {\partial q_\alpha} = - \frac{g_s}{A 2 \pi} \text{Im}
\int\!\! dE\, n_F(E)  \sum_{\bm k} \\\notag
\text{Tr}\{ 
\hat{\cal G}_{\bm k}(E)
\hat{\gamma}^{\alpha}_{\bm k}
\hat{\cal G}_{\bm k}(E)
( \hat{\cal G}_{\bm k - \bm q}(E) - \hat{\cal G}_{\bm k + \bm q} (E))
\}\;.
\end{align}
We observe a clear anisotropy with a pronounced spike that hints at a
singular behavior for the results in the {\it y} direction, absent in
the {\it x} direction, where the behavior is closer to that of the
(analytically known) Dirac cone approximation \cite{supplmaterial}. The
numerical results strongly suggest the restoration of a regular {\it
  2d} response but with strong anisotropy. This is confirmed by the
analytical treatment that follows.

Now we obtain analytically the charge response to lowest order in lattice
effects. We start with the determination of the prefactor $f_{+}(\bm
q)$, that is, Eq. (\ref{prefactor}) for two states on opposite sides
of the Fermi surface: $\bm k_2$ and $\bm k_1 = \bm k_2+ \bm q$, such
that the vector $\bm q$ corresponds to the square-root singularity in
the response.  The latter condition requiring the $\bm q$-linked
portions of the Fermi surface to be parallel. Upon a Jacobi-Anger
expansion of the terms \mbox{$\exp(i\bm k \cdot \bm \delta_i)$}, the
structure factor can be written as $S(\bm k) = 3 \sum_{n} J_{-1+3n}(k'a)
e^{i(-1+3n) \phi}$ where $J_n$ are Bessel functions of the first kind,
and the separation from the Dirac point is $\bm k' = \bm k - \bm K_1$,
with polar coordinates $(k',\phi)$. To lowest order in lattice
effects, only $J_{-1} = -J_1$ and $J_2$ are to be retained. Then, the
requirement $|S(\bm k)| = \text{const}$ leads to the following
expression for the Fermi surface in polar coordinates:
\begin{equation}\label{fermisurface}
k'_F(\phi) = k_F \left( 1 + \frac{k_F a}{4} \; \cos(3 \phi) + 
{\cal O}(k_F a)^2\right)  
,\end{equation} %
where we have parametrized the Fermi energy by the would-be Fermi wave vector
in the isotropic limit: $ E_F = \hbar v k_F$.

To lowest order, the condition of parallel pieces of Fermi surface
leads to the following relation between polar angles $(\phi_{1,2})$ of
the involved $\k$-points: $\phi_1 = \phi_q - \delta \phi$ and $\phi_2 =
\phi_q + \pi + \delta \phi$ with the lattice correction $\delta\phi =
\frac{3}{4} \; (k_F a) \; \sin(3 \phi_q)$ and $\phi_q$ the polar
angle of the joining vector $\bm q$ with modulus $q = 2 k_F
( 1 + {\cal O} (k_F a)^2)$. We can now write the phase of $S(\bm k')$ as 
\begin{equation}  
\frac{S(\bm k')}{|S(\bm k')|} \sim - e^{-i (\phi + \theta)} 
,\end{equation} %  % 
where $\phi$ is the polar angle of $\bm k'$ and $\theta$ is the lattice correction to that phase given to lowest order by $\theta =  \frac{1}{4} (k_F a) \; \sin(3\phi)$. This leads to the final expression for the prefactor
\begin{equation}
\label{fplus}   
f_{+}(\bm q) = \frac{(k_F a)^2}{8} (1 - \cos(6 \phi_q)) + {\cal O}
(k_F a)^3\; ,
\end{equation} 
where $(2 k_F, \phi_q)$ are the polar
coordinates of $\bm q$ and which holds for both Dirac points. We note that the
result of Eq. (\ref{fplus}), although the lowest finite order in a $k_F$
expansion, already represents an excellent approximation for sizable Fermi
levels well within the range of gate-voltage doped graphene's samples \cite{supplmaterial}.

Combining Eqs.  (\ref{chirhofake}) and (\ref{fplus}), the
dominant singularity of the {\em true} response is
\begin{equation}\label{chirhosingular}
\Pi \sim  \frac{(k_F a)^2}{8} (1 - \cos(6 \phi_q))\frac{g_s g_v \sqrt{k_F}}{\pi \hbar v} 
\sqrt{q - 2 k_F} \; \Theta(q - 2 k_F)\;.
\end{equation}
We can now determine the density response associated to a local
perturbing potential $V = u\delta(\bm r)$ given by $\frac{\delta\rho(\bm r)}{u} = \frac{1}{(2 \pi)^2} \int d^2 q \; e^{i \bm q \cdot \bm r} \Pi(\bm q)$.
The remaining integral is obtained from
standard techniques \cite{lighthill58}, leading to the following asymptotic
behavior for Friedel oscillations:
\begin{equation}
\label{analytic}
\frac{\delta\rho(\bm r)}{\rho_e} = -\frac{(k_F a)^2}{2 \sqrt{2}} (1-\cos(6 \phi_r))
\frac{u\rho_e}{E_F}
 \frac{\sin(2 k_F r)}{(2 k_F r)^2}\;.
\end{equation}
While its $r^{-2}$ behavior is standard for a {\it 2d} system,  its true origin as a lattice contribution to an otherwise {\em null} result (to order $r^{-2} $) is revealed by its amplitude, vanishing as $ (k_F a)^2$ in the scaling limit $a\rightarrow 0$, and by its anisotropy reflecting the sixfold symmetry of
the lattice.

Comparing Eq. (\ref{analytic}) with the result coming from the Dirac cone
approximation, Eq. (\ref{FOAnalyticDirac}), we first notice the phase shift of
$\pi/2$. We further find for the crossover length between anomalous and regular
Friedel oscillations $r_c\sim k_F^{-1}(k_Fa)^{-2}$. For $k_F=1$nm$^{-1}$, we
have $r_c\sim100$nm which corresponds to an impurity concentration of
$n_i\sim10^{10}$cm$^{-2}$, recently found to be the intrinsic concentration of
local impurities in graphene \cite{Ni10}. We thus expect Friedel oscillations to
show anisotropic behavior and modify the RKKY-interactions for doping levels
$E_F\gsim0.5$eV.

%%%%%%%%%%%%%%%%%%%%%%%%%%%%%%%%%%%%%%%%%%%%%%%%%%%%%%%%%%%%%
%  SECTION SUMMARY
%%%%%%%%%%%%%%%%%%%%%%%%%%%%%%%%%%%%%%%%%%%%%%%%%%%%%%%%%%%%%

{\it Summary.} The simplest tight-binding model has been employed to study the
lattice contribution to the magnetic susceptibility  and (coarse-grained) charge
response of doped graphene, for which the Dirac cone approximation produces a null
result. The lattice magnetic response shows orbital paramagnetism for a wide range of
filling factors, representing a relevant contribution when compared to other sources
such as core diamagnetism, spin paramagnetism and  electron-electron interaction
induced orbital paramagnetism. Lattice effects restore the {\it 2d} regular behavior
for the coarse-grained charge response,   with  Friedel oscillations decaying as
$r^{-2}$ but with pronounced sixfold anisotropy, with maxima along bond's directions.
For clean samples with impurity concentrations of   $n_i\sim10^{10}$cm$^{-2}$ they
become relevant for doping levels    $E_F\gsim0.5$eV.

{\it Acknowledgments.} We are grateful to  F. Guinea for useful
discussions. This work has been supported by FCT under grant PTDC/FIS/101434/2008 and MIC under grant FIS2010-21883-C02-02. 

%\bibliography{suscargaetal}

\providecommand{\noopsort}[1]{}\providecommand{\singleletter}[1]{#1}%

\end{document}